# Scanning Transmission Electron Tomography and Electron Energy Loss Spectroscopy of Silicon Metalattices

*Shih-Ying Yu[a], Hiu Yan Cheng[b], Jennifer L. Dysart[b], ZhaoHui Huang[c,d], Ke Wang[c], Thomas E. Mallouk[b], Vincent H. Crespi[a,b,c,d], John V. Badding[a,b†], and Suzanne E. Mohney[a,c*]*

[a] Department of Materials Science and Engineering, The Pennsylvania State University, University Park, Pennsylvania, 16802, United States of America

[b] Department of Chemistry, The Pennsylvania State University, University Park, Pennsylvania, 16802, United States of America

[c] Materials Research Institute, The Pennsylvania State University, University Park, Pennsylvania 16802, United States of America

[d] Department of Physics, The Pennsylvania State University, University Park, Pennsylvania 16802, United States of America

[†] Deceased October 26, 2019.

Abstract: Transmission electron microscopy, scanning transmission electron tomography, and electron energy loss spectroscopy were used to characterize three-dimensional artificial Si nanostructures called "metalattices," focusing on Si metalattices synthesized by high-pressure confined chemical vapor deposition in 30-nm colloidal silica templates with ~7 and ~12 nm "meta-atoms" and ~2 nm "meta-bonds". The "meta-atoms" closely replicate the shape of the tetrahedral and octahedral interstitial sites of the face-entered cubic colloidal silica template. Composed of either amorphous or nanocrystalline silicon, the metalattice exhibits long-range order and interconnectivity in two-dimensional micrographs and three-dimensional reconstructions. Electron energy loss spectroscopy provides information on local electronic structure. The Si $L_{2,3}$ core-loss edge is blue-shifted compared to the onset for bulk Si, with the meta-bonds displaying a larger shift (0.55 eV) than the two types of meta-atoms (0.30 and 0.17 eV). Local density of state calculations using an empirical tight binding method are in reasonable agreement.

**Keywords:** Silicon, chemical vapor deposition, metalattice, electron tomography, electron energy loss spectroscopy, quantum confinement

When the features of semiconductors are nanoscale, many of their intrinsic properties change dramatically, offering opportunities for improved or new functionality for nanoelectronics,[1] photonics,[2] thermoelectrics,[3,4] and spintronics.[5] Three-dimensional (3D) ordered materials synthesized in templates such as polystyrene (PS) or silica spheres have been successfully used to create photonic bandgap materials.[6,7] While photonic crystals were initially studied on the scale of optical wavelengths, theorists[8–10] have since considered nanoscale structures at the length scale of the wavelength of electrons, phonons, and excitons. Experimentally, preparing such structures is more challenging than creating the more widely studied photonic bandgap materials. As pores in the templates used to create these materials shrink to a few nanometers, conventional chemical vapor deposition (CVD) fails to effectively transport the precursors into the openings.[11]

Recently, high-pressure confined chemical vapor deposition (HPcCVD) has attracted attention because of its ability to conformally deposit semiconductors in high aspect ratio optical fibers. Semiconductors including Si,[12–14] Ge,[15,16] and ZnSe[17] have been successfully deposited in centimeter-long fibers without forming voids. Semiconductor metalattices—a new artificial nanostructure that can be viewed as an interconnected 3D network of nanocrystals (meta-atoms) and nanowires (meta-bonds)—have exhibited interesting thermal[18] and mechanical[19] properties that differ markedly from bulk Si. They are also expected to demonstrate unique electronic structures due to quantum confinement of the smallest features and proximity of adjacent meta-atoms and meta-bonds.

Although conventional transmission electron microscopy (TEM) provides nanoscale information in two directions perpendicular to the transmitted beam, the projection nature of TEM limits the interpretation of information along the third dimension. Electron tomography is a 3D technique that can reconstruct 3D structures based on a series of 2D projections and is used in this work. While routine in biological sciences, electron tomography began to be applied in materials science only with the development of Z-contrast tomography.[20,21] Examples of 3D characterization of nanoporous metals,[22,23] zigzag nanowires,[24] and porous thin films[25,26] show the great advantage of this technique to reveal both external and internal structures of the observed objects. Possible misinterpretations of 2D projection images can be reduced if tomography results are provided.[27]

The metalattices we imaged are especially interesting because the electronic and optical properties of Si are modified in sufficiently small Si nanocrystals,[28–32] Si nanowires,[33] Si/SiO$_2$ multilayer superlattices,[34,35] or porous silicon.[36,37] In the metalattice, quantum confinement of the smallest features and their proximity lead to a unique electronic structure. In this work, spatially resolved electron energy-loss spectroscopy (EELS) is used to track shifts in the Si L$_{2,3}$ core-loss edge as a function of position in the lattice. Local density of state calculations using an empirical tight binding method are also performed to correlate the local degree of confinement in different regions of the structure and analyze variations along real-space paths in the structure.

Illustrations of a three-step process for fabricating a Si metallatice are shown in Fig.1 (a) with the structure in more detail in Fig. 1 (b). Field emission scanning electron microscopy (FESEM) of the steps are shown in Figs. 1 (c–e). Based on previous literature, both experimental[38] and theoretical[39], it has been shown that FCC is the more stable close-packed stacking for the spheres in the template compared to the hexagonal close-packed (HCP) structure. A fast Fourier transform (FFT) of a metallatice showing a {110}-type plane of the FCC structure is shown in Fig. S1 and cannot be indexed to any of the HCP planes.

After thin cross-sections were prepared using a focused ion beam (FIB) (Supporting Information), an FEI Titan3™ G2 60-300 S/TEM and FEI Talos operated at 200 kV were used for TEM. Selective area diffraction (SAD), bright field TEM (BF-TEM), and high angle annual dark field (HAADF) images were acquired in scanning transmission electron microscopy (STEM) mode.

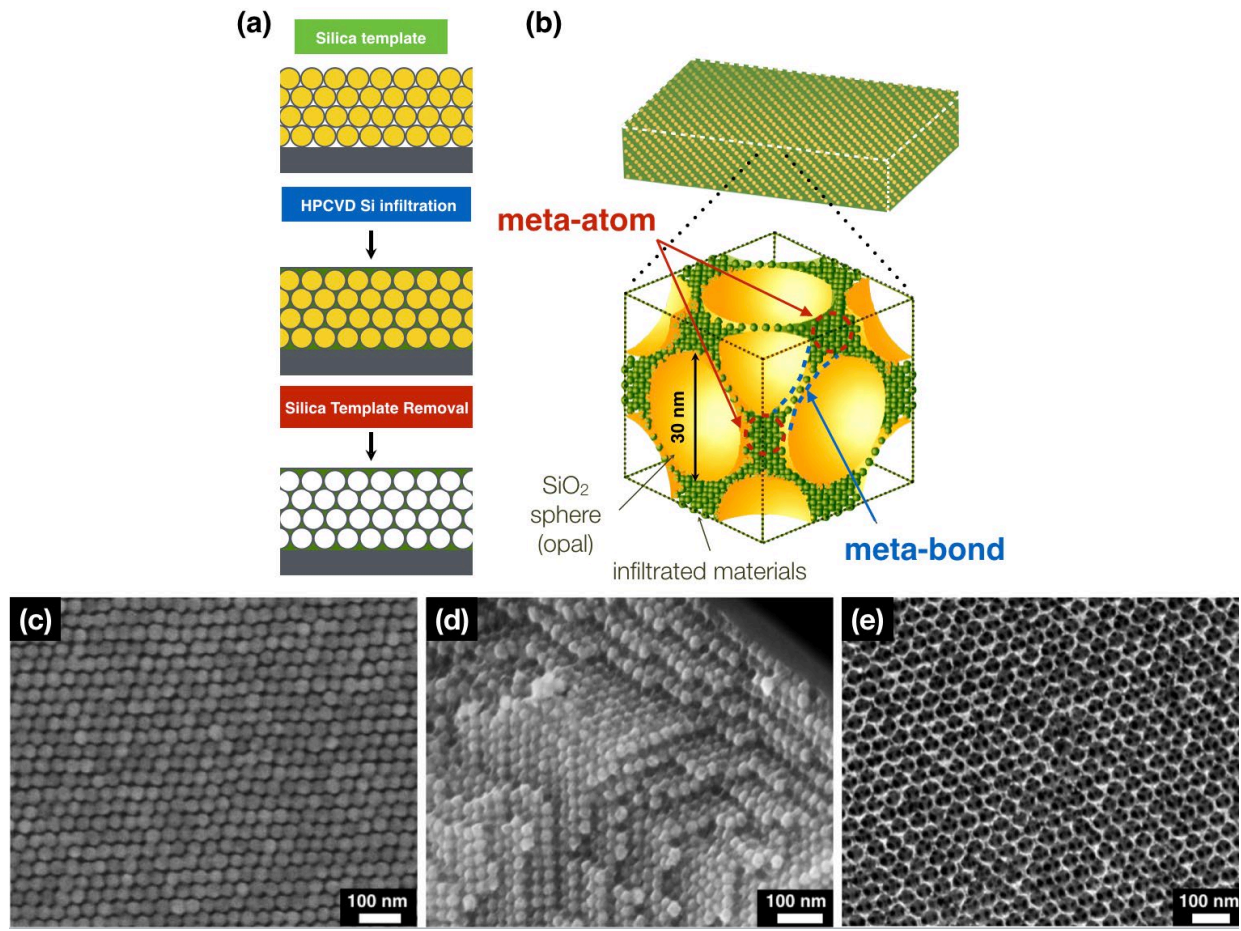

**Figure 1** (a) Close-packed silica spheres (yellow) were filled with Si (green) using HPcCVD and then removed, leaving only the Si metalattice. (b) Interconnected clusters (meta-atoms) and nanowires (meta-bonds) made of Si atoms (depicted as green spheres) are labeled. (c–e) FESEM of the three steps from (a): (c) top view of close-packed silica spheres (template), (d) fracture surface of a filled template, and (e) top view of a Si metalattice with template removed.

Figure 2 (a) is BF-TEM showing a cross sectional view of the sample. The as-deposited Si is amorphous (Fig. S2) and can be crystallized by annealing. The SAD pattern in the inset of Fig. 2 (a) reveals that annealed Si is nanocrystalline without preferred orientation. From the HAADF image at higher magnification in Fig. 2 (b), we can see the meta-atoms at octahedral and tetrahedral interstitial sites of the FCC template, connected to each other by meta-bonds. Energy dispersive spectroscopy (EDS) mapping shown in Fig. 2 (c) confirmed that the metalattice is pure Si within the detection limit of the instrument.

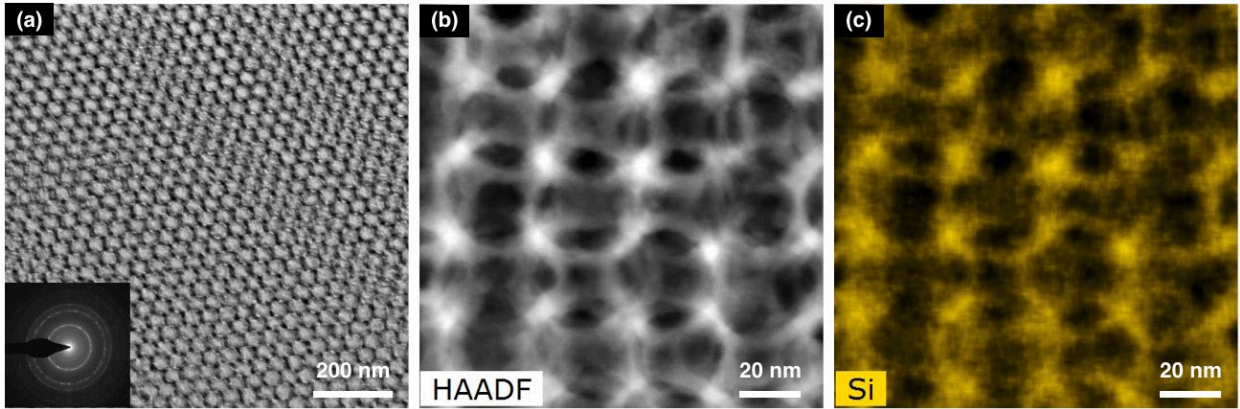

**Figure 2** (a) Bright field TEM image of the metalattice. The inset is the selected area diffraction (SAD) pattern of the infiltrated silicon after annealing. (b) High angle annual dark field (HAADF) image and (c) EDS mapping showing the distribution of Si.

Typical TEM samples may include low-dimensional structures that are uniform along the third dimension. If so, 2D images are readily interpreted. For our periodic metalattice with ~30 nm repeat distances, the resulting contrast in TEM images includes a contribution from overlapping signals of each component at different heights as the electron beam goes through the sample. Some features might actually be very different from what appears in a 2D projection. Since the sizes and morphologies of nanomaterials are key to understanding and, therefore, adjusting their properties, it becomes imperative to combine both conventional TEM (2D) and electron tomography (3D) to give a full picture.

To obtain high-fidelity tomography results, the signals of the 2D images need to fulfill the projection requirement.[40] Therefore, the tomography experiment was carried out in HAADF-STEM mode in the FEI Titan3™ with details on the holder and tilt series in the supporting information. The intensity of an atom in HAADF imaging scales with atomic number ($Z^{~1.7}$) due to Rutherford scattering, so it is suitable for crystalline materials. Single-axis tilt series were acquired automatically using the tomography module in the FEI Xplore3D software suite.

Data sets were acquired over a total tilt range of 65° to -65°; increments of 1° in the low angle region (55° to -55°) and 0.5° in the high angle region (65 to 55° and -65° to -55°) were set to enhance the reconstruction resolution. At the tilting angle 1°, the FFT of the HAADF image in Fig. 3 (a) showed that the {114} plane was perpendicular to the transmitted beam. As the sample was negatively tilted, another low index zone axis appeared at -34°. The FFT showed a six-fold symmetry, which indicated the approach of a <111> zone axis. Assuming we started from the [114] zone axis, the sample required a 35.26° tilt to arrive at the [111] zone axis, and the angle difference between these two projections (1° and -34°) is just the difference between [114] and [111]. The angular relationships between the above zone axes are depicted on the Wulff net in Fig. 3 (b). We aligned the tilt axis along a line connecting the north and south poles and put the [114] zone at the center for convenience. Based on the drawing, we expected to meet the [332] zone axis at 45.29° with respect to [114], i.e., 10.03° from [111]. However, the HAADF image and its FFT acquired at -41° is the closest one to [332] with about 2° difference between the measured pattern and standard pattern. Considering that we only tilted along a single direction and the measurement was made on the FFT of the image at high-tilt angle, it is possible that this zone axis was not on the tilting trail and a few degrees off is expected. Fortunately, the relatively large lattice has a larger error allowance in orientations compared to atomic lattices such that we can still observe the desired planes even if we are not right on the zone axis. We simulated several FFTs from the plane that was observed slightly away from the [332] zone axis and found that our measured FFTs matched the simulated FFT for 4° off the [332] zone axis. Therefore, by looking at the sample at multiple angles, we confirmed the conclusion we made from the 2D TEM images that our template, as well as the Si metalattice, are FCC structures.

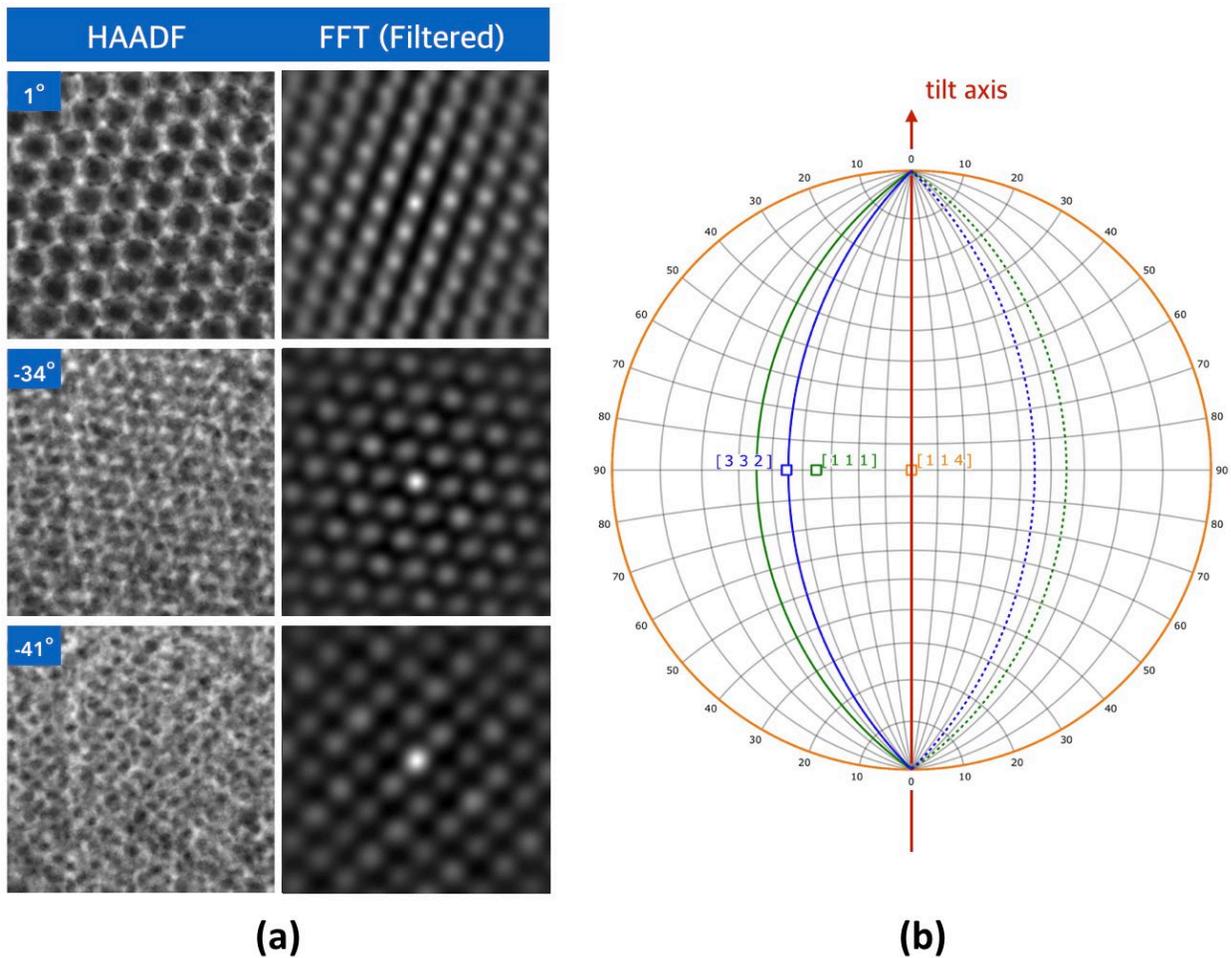

**Figure 3** (a) The HAADF images and the corresponding FFTs titled 1°, -34° and -41°. (b) Trail of tilting series during acquisition of 2D projections is represented by putting the [114] zone at the center of the projection. The stereogram is constructed on the Wulff net to show the angular relationships between different zone axes.

The procedure for reconstruction is provided in the supporting information, and an example of orthogonal cuttings of the reconstruction is shown in Fig. 4 (a). The continuity between different planes was examined to ensure that a correct 3D structure was established. Figure 4 (b) is the volume rendering from the reconstructed results, offering an overall impression of 3D concentration distribution. Further details about the reconstruction is supplied as Supporting Information, although an animation is not available through arXiv.org. The 3D model shows that the metalattice is highly porous and interconnected. The formation of small necks between silica spheres occurred during sintering to add strength to the templates before infiltration, resulting in interpores (yellow arrows in Fig. 4 (b)). The diameters of the necks are controlled by the sintering temperature and time, providing flexibility for modifying the fine structure of the metalattice. The average diameter of the interpores is ~3.5 nm. The volume fraction of the silica for FCC stacking of the spheres is 74%, provided there are only point contacts between spheres. As spheres in the template start to overlap more, the porosity of a resulting metalattice further increases. This information is especially important when considering transport properties, which researchers expect to be affected by porosity and distribution of pores.[41-43]

A common phenomenon reported for larger photonic crystals deposited by atomic layer deposition (ALD) was that they were not completely void-free.[44–46] Due to the highly conformal growth characteristics of ALD, there is a geometrical limit for fully filling all the voids inside this type of template. The diffusion channels close as the smallest pores in the {111} planes in the template are filled. Thus, the maximum infiltration by conformal filing of an FCC opal crystal is 86% of the empty space (22 % of total space), leaving the interstitial pores partially filled. Voids can usually be observed at the center of octahedral and tetrahedral sites since those are two of the relatively large spaces in the template. Examples of void formation[45,46] illustrate the geometric limit for filling a close-packed template.

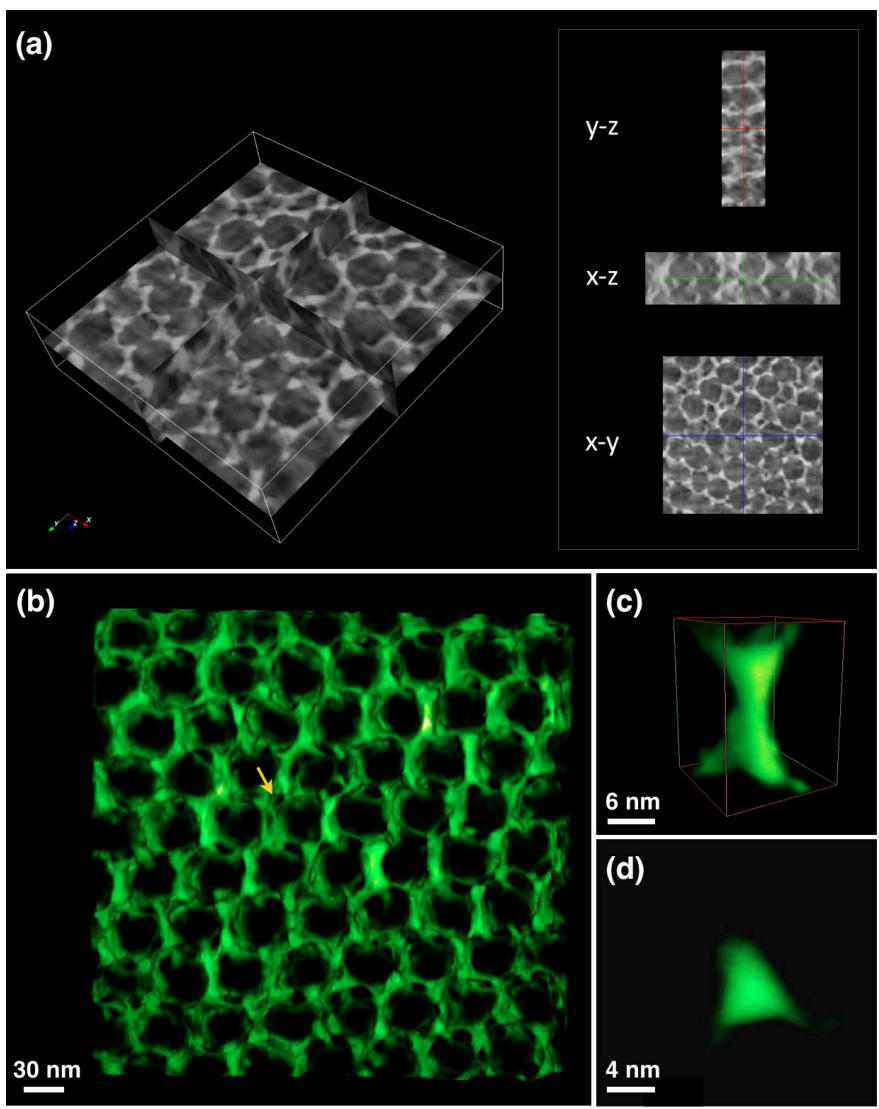

**Figure 4** (a) Orthogonal slices from the reconstructed 3D structure. The two-dimensional image of the y-z, x-z and x-y planes are shown separately on the right. (b) Volume rendering of Si metalattices and (c) octahedral and (d) tetrahedral interstitial sites cropped from the metalattices. The yellow arrow points to an "interpore".

We dissected the metalattice prepared by HPcCVD into two features and represented meta-atoms at octahedral and tetrahedral sites in the silica template, as shown in Fig. 4 (c) and (d), respectively. The resulting sizes of the octahedral and tetrahedral interstitials sites are ~70-80% of those expected from an ideal FCC template. This effect can be attributed to either volume loss during sintering of the template or over-etching of the silica after HPcCVD. However, the shapes of the interstitial sites are still very similar to those of the interstitial sites from the FCC silica lattice. The orthogonal slices revealed that there are no obvious pores in the Si meta-atoms at the resolution of electron tomography (~1 nm). Compared to ALD, which by calculation could leave voids with diameters of 7 nm and 2 nm in filled templates composed of 30 nm spheres, HPCVD proved to be an excellent technique for fully filling the close-packed template.

Electron energy loss spectroscopy (EELS) provided local electronic structure information due to its high spatial and energy resolution. A Gatan Quantum spectrometer was used to obtain the EELS and HAADF-STEM images for this study. In order to obtain atomic resolution STEM images, both image and probe corrector were carefully calibrated before characterization. A monochromator was used to ensure the best energy resolution. Figures 5 (a–d) are HAADF images of sites in the sample that were probed, and Fig. 5 (e) is the Si $L_{2,3}$ edge for the Si reference and three different features of the Si metalattice. The energy resolution was determined from the full width at half maximum of the zero-loss peak (not shown) and was calibrated frequently during the experiment to maintain it in the range 0.15–0.20 eV. Here we used the silicon substrate underneath the Si metalattice film as the bulk reference at 99.5 eV, as shown in Fig. 5 (e). As we moved the electron beam to parts of the metalattice, as shown in Fig. 5 (b–d), there was a blue shift of 0.17 and 0.30 eV for the two types of meta-atoms, and it reached a maximum of 0.55 eV at the meta-bond. Moreover, the shape of the $L_{2,3}$ edge also changed compared to that of bulk Si, and it is similar to the results observed for Si nanocrystals with diameters below 5 nm.[30] Based on the observation of the Si core-loss edge, we conclude that quantum confinement alters the band structure of the Si metalattice.

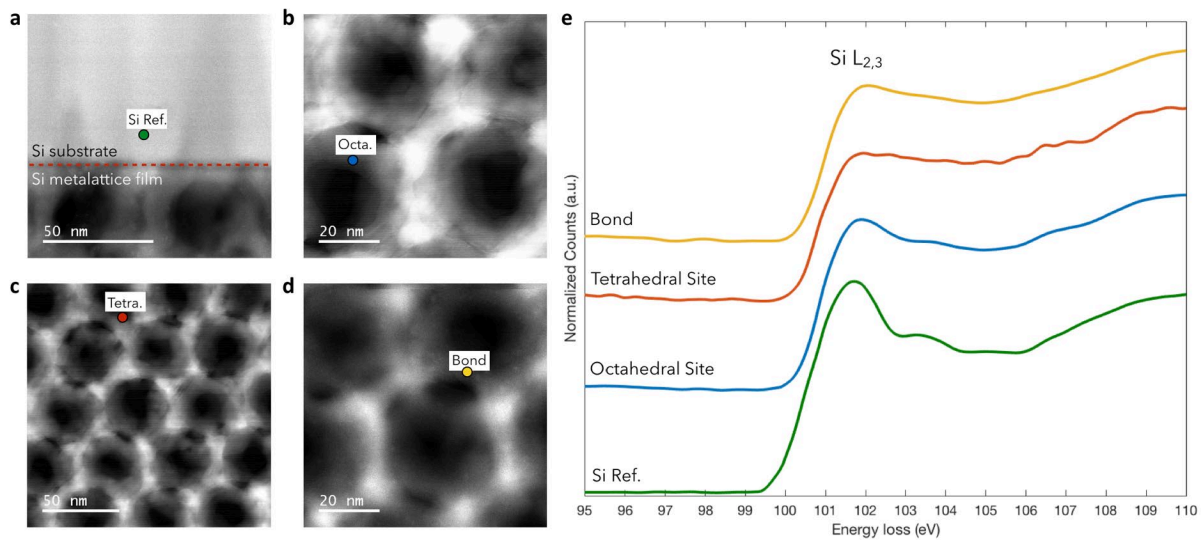

**Figure 5** STEM-EELS of the silicon metalattices probed at different regions. (a) High-angle annular dark-field (HAADF) image taken at 200 kV near the interface between the Si substrate and the Si metalattice film. (b-d) HAADF images of meta-atoms at octahedral site, tetrahedral site, and meta-bond. (e) Normalized Si L2,3 EELS core-loss spectra corresponding to the selected areas indicated in a-d. All of the spectra measured from the metalattice showed the edge shape changes and shifts upward in energy as feature size decreases.

To model these alterations, we constructed a unit cell for the 30 nm silicon metalattice consisting of over 100,000 silicon atoms, which emphasizes low-energy surfaces that are then further terminated in the simulation with hydrogen to remove dangling bonds (Fig. 6 (a, b)). The relative dimensions of the meta-atoms and meta-bonds is constrained by setting the total silicon volume fraction to equal that of the void space in a close-packed sphere template and also maintaining the octahedral/tetrahedral volume ratio of this template. The local density of states as a function of energy is calculated by an empirical tight-binding method that exploits matrix sparsity to handle large systems, averaging the density of states over localized regions that follow high-symmetry lines connecting meta-atoms to meta-bonds. (See Supporting Information.) These results reveal a continuous variation in the conduction band edge along the real-space path from tetrahedral site to meta-bond to octahedral site in Fig. 6 (c, d), with the bandgap varying in a manner correlated to the local degree of confinement in different regions of the structure. The measured EELS blue shifts are in reasonable agreement with those from the calculations and are robust across multiple measurements.

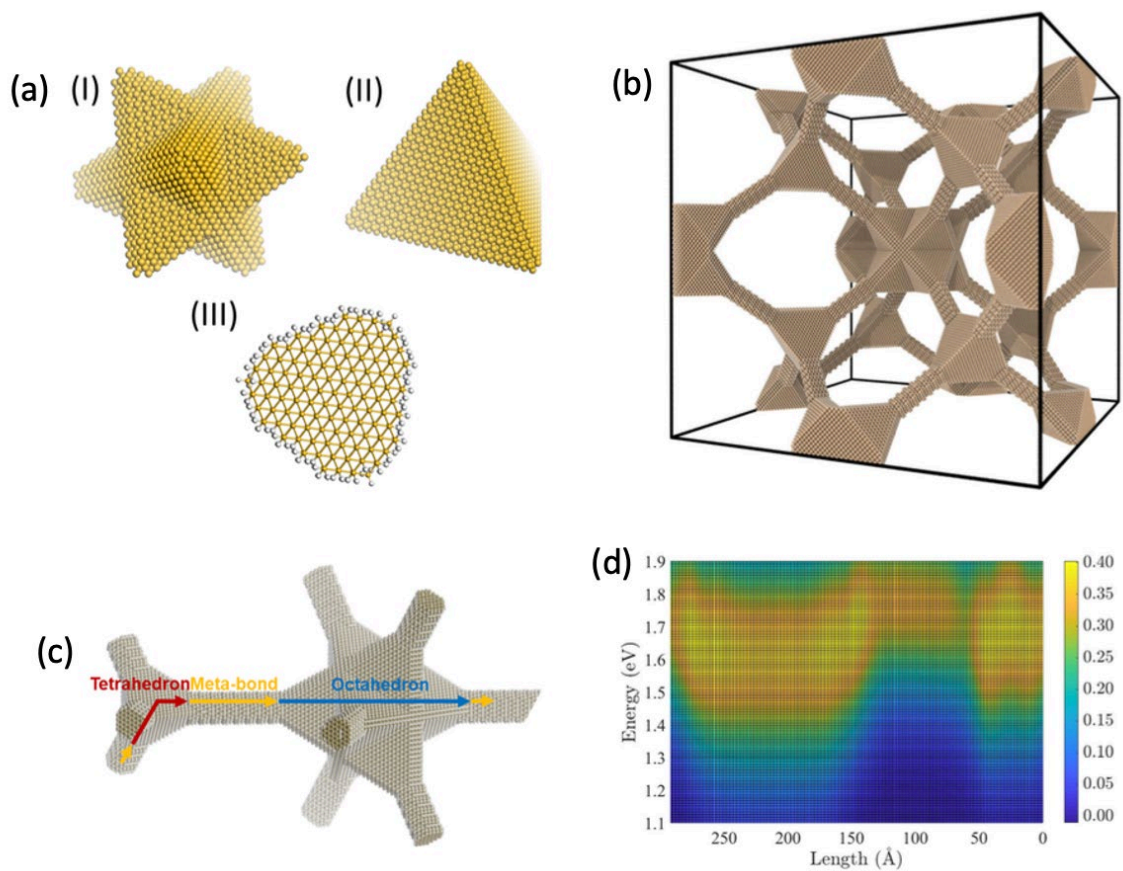

**Figure 6 (a)** (I) Stellated octahedron and (II) tetrahedron sites used for unit cell construction, and (III) meta-bond cross section showing hydrogen termination for a low free-energy surface, (b) unit cell construction of metalattice for tight-binding calculations, (c) single unit cell showing calculation paths traveling in real space from tetrahedral site to metal bond to octahedral site, and (d) empirical electronic structure calculation of Si metalattice in real space, showing the local density of states projected onto a region whose center travels along the high-symmetry line depicted in (c); this shows the modulation in the conduction band position as the degree of quantum confinement varies along the path. The conduction band minimum follows essentially the lowest-energy orange/yellow coloration in the plot.

The silicon in the metalattice is polycrystalline. Therefore, we also assessed the impact of grain boundaries in modulating the effects of confinement by examining silicon structures bounded by model Σ3 grain boundaries and found that they do not greatly impact the positions of the band edges, suggesting that the polycrystallinity of the metalattice silicon does not obscure the spatial modulation of the quantum confinement in this extended structure. More detail is provided in the Supporting Information, including Fig. S3.

In summary, this work reveals the real space and local electronic structure of a three-dimensional Si metalatice prepared by HPcCVD. The metalattice is well ordered and displays quantum confinement due to the small size of the meta-atoms and meta-bonds. With further work on nanoscale metalattices of different sizes prepared from a variety of semiconductors, the opportunity to engineer unique electronic, thermal, mechanical, and optical behavior of semiconductor metalattices is ripe for exploration.

ASSOCIATED CONTENT

**Supporting Information**. Provided information include: FESEM micrograph showing the metalattice and its fast Fourier transform (Fig. S1), details on sectioning samples, TEM micrograph showing the metalattice and diffraction from infiltrated amorphous Si (Fig. S2), information on the electron tomography holder and more on the tilt series and Wulff net, and further information on the electronic structure calculations with Fig. S3.

Supporting Information is appended.

AUTHOR INFORMATION

**Corresponding Author**

*Email: mohney@psu.edu

**Author Contributions**

Dr. Shih-Ying Yu performed the microscopy in this paper and received assistance during some sessions from Dr. Ke Wang. Portions of this work are from Dr. Yu's Ph.D. thesis. Dr. Hiu Yan Cheng performed HPcCVD to deposit Si in templates that Dr. Jennifer L. Russell prepared. Dr. ZhaoHui Yang performed local density of states calculations, and his advisor Prof. Vincent H. Crespi wrote about those results. Drs. Jennifer L. Russell, Hiu Yan Cheng, and Shih-Ying Yu were advised as students by Profs. Thomas E. Mallouk, John V. Badding, and Suzanne E. Mohney. The authors have given approval to the final version of the manuscript.

ACKNOWLEDGEMENTS

The authors acknowledge the financial support from the Penn State Materials Research Science and Engineering Center through NSF award DMR 1420620. S. Yu would also like to thank Dr. Jenn Gray for helpful discussions on electron tomography.

# Scanning Transmission Electron Tomography and Electron Energy Loss Spectroscopy of Silicon Metalattices

Shih-Ying Yu[a], Hiu Yan Cheng[b], Jennifer L. Dysart[b], ZhaoHui Huang[c,d], Ke Wang[c], Thomas E. Mallouk[b], Vincent H. Crespi[a,b,c,d], John V. Badding[a,b†], and Suzanne E. Mohney[a,c]*

[a] Department of Materials Science and Engineering, The Pennsylvania State University, University Park, Pennsylvania, 16802, United States of America

[b] Department of Chemistry, The Pennsylvania State University, University Park, Pennsylvania, 16802, United States of America

[c] Materials Research Institute, The Pennsylvania State University, University Park, Pennsylvania 16802, United States of America

[d] Materials Research Institute, The Pennsylvania State University, University Park, Pennsylvania 16802, United States of America

**Metalattice Synthesis**

To prepare the metalattices, monodisperse silica sphere colloids with a diameter of 30 nm were synthesized by hydrolysis of tetraethylorthosilicate, which was slowly released from a cyclohexane layer above an aqueous reaction medium, in the presence of L-arginine.[1] The resulting colloids were dispersed in ultrapure deionized (DI) water. A silicon substrate was then immersed into this colloidal solution and heated in a well-insulated, humidity-controlled box furnace for two weeks. Silica spheres were self-assembled into the close-packed template by using a vertical deposition technique.[2] The resulting dried film was pre-sintered at 650 °C for 1 h to fuse neighboring spheres and improve internal connections. The as-synthesized template film was approximately 100 nm to 1 μm thick, depending on the growth parameters. This template was then infiltrated with amorphous Si by HPcCVD. The high-pressure deposition was performed in a stainless-steel reactor, using a gas mixture of $SiH_4$ and He with pressures up to 35 MPa and a deposition temperature of 350–400 °C The as-deposited Si metalattice was then annealed at 800 °C to crystallize the infiltrated material. The sacrificial silica template was removed by selective etching using 1:10 buffered oxide etch (BOE), leaving only the Si metalattice for characterization.

**Field Emission Scanning Electron Microscopy**

The morphologies of the template were initially investigated using a Leo 1530 field emission scanning electron microscope (FESEM). A fast Fourier transform (FFT) is consistent with the {110} FCC structure and cannot be indexed as HCP (Fig. S1).

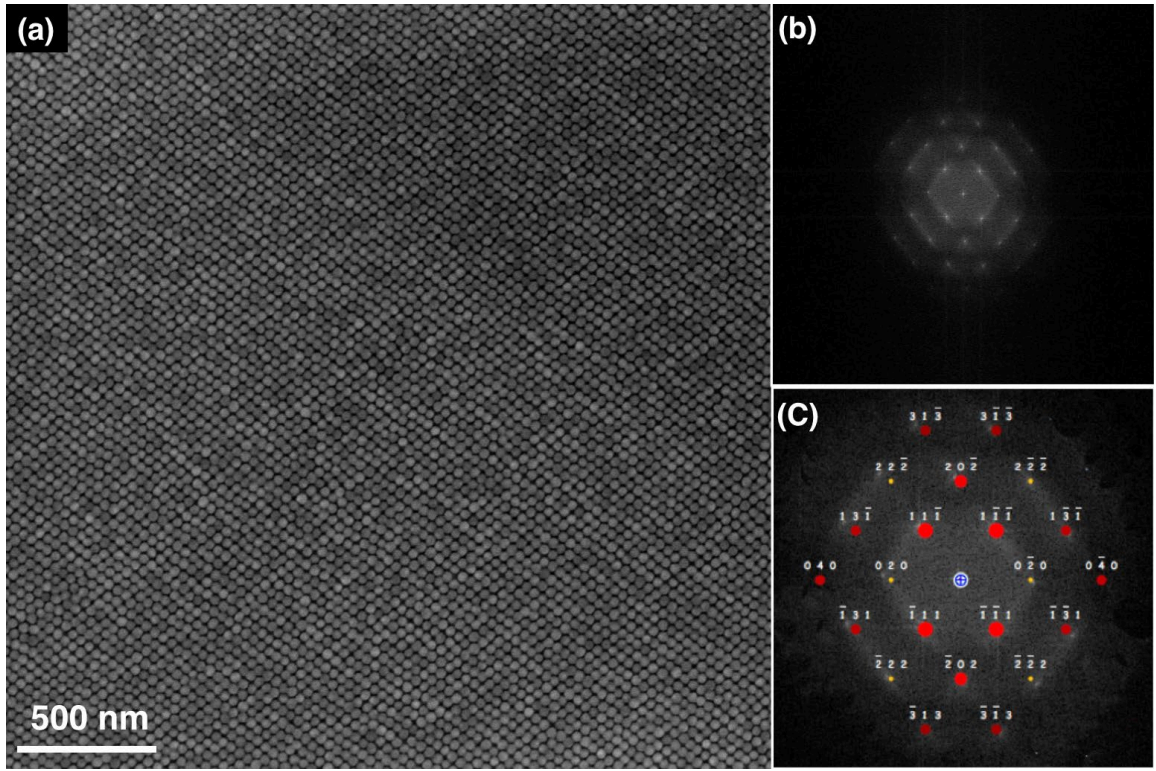

**Figure S1** (a) Top-view SEM images of a Si metalattice and its (b) FFT and (c) the indexing results of the FFT, showing it is oriented with a {110} type plane.

**Sectioning for Transmission Electron Microscopy**

Thin sections for TEM were prepared by a FEI Helios NanoLab™ DualBeam™ 660 system. A protective carbon layer was first deposited on top of the metalattice, and an in-situ lift-out technique was used to transfer the lamella from the sample to a TEM grid. Damage from $Ga^+$ ions and surface amorphization were reduced by using a final 2 kV cleaning procedure. In order to preserve at least a full unit cell of the metalattice, the total thickness of the film was controlled from 60–100 nm during focused ion beam (FIB) sample preparation.

**Microscopy of an Amorphous Silicon Metalattice**

Before annealing, the Si infiltrated into the metalattice was amorphous (Fig. S2).

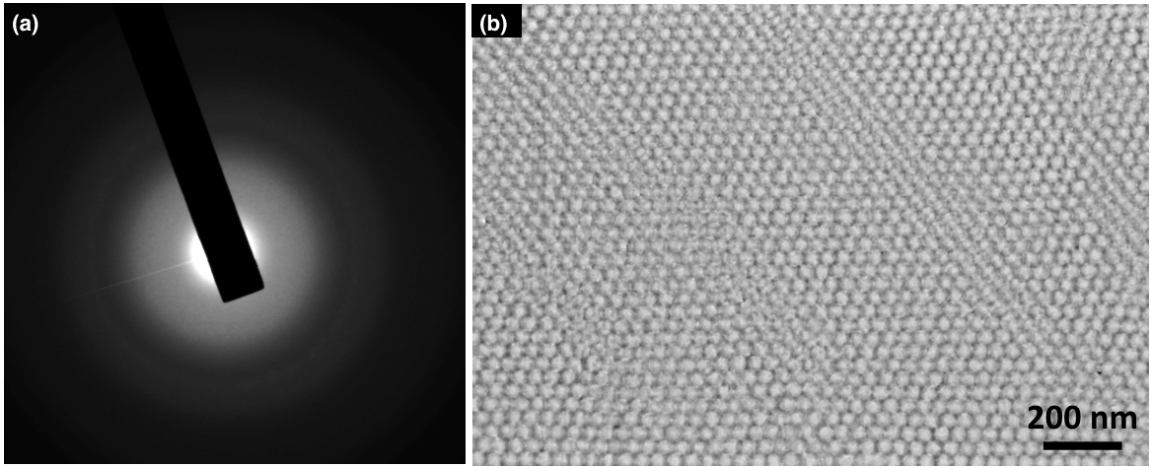

**Figure S2** (a) SAD and (b) BF-TEM image of as-deposited Si metalattice, confirming that the Si is amorphous after HPCVD infiltration.

**Electronic Tomography Holder and Tilt Series**

A series of images at different angles was recorded for analysis and further reconstruction. A Fischione advanced tomography holder (model 2020) was used to eliminate shadowing effects encountered by most common TEM holders at high-tilt angles. A small convergence angle was chosen (<10 mrad) to enlarge the depth of focus of imaging to prevent the object being out of focus at high tilt angles. During tilting, the image shift relative to the reference image was corrected using a cross-correlation method, and an autofocus routine was performed using a through-focus series before taking each image. A 3D reconstruction was carried out using a simultaneous iterative reconstructions technique (SIRT) in FEI Inspect3D reconstruction software. Visualizations of the 3D tomography reconstruction results was performed using FEI Avizo® 8 3D visualization software.

**3D Reconstruction**

Each 2D projection obtained at different tilting angles has a dimension of 2048×2048 pixels, representing 335×335 nm. A sequential post-alignment of images was done by using a cross-correlation algorithm before 3D reconstruction. Since the cross-correlation algorithm is not sensitive to the small change of the tilt axis orientation, an additional alignment of the tilt axis is achieved by using the feature tracking function built into the Inspect3D software. This process was repeated until the smallest deviation of feature position during tilting was achieved. For our data, the average deviation is about 0.5 nm after 5 iterations.

**Electronic Structure Calculations**

Due to the large system size, the Implicitly Restarting Arnoldi Method (IRAM),[3] a derivative of the primitive power method, is applied to iteratively find an energy window, i.e., a Krylov invariant subspace, whose eigenpairs are used to approximate the eigenpairs of the original Hamiltonian by the Rayleigh-Ritz procedure. A spectral shift and invert strategy is also applied to obtain only a small subset of eigenpairs. The energy window can then be scanned and covers the full eigenspectrum of interest. IRAM is implemented in a parallel tight-binding code. To prevent the Krylov subspace from becoming too large, an implicitly restarting algorithm is used to deflate the subspace by removing eigenpairs that are far from the region of interest and then resuming the creation of the vector power sequence from a smaller upper Hessenberg matrix. The IRAM algorithm does not store the whole Hamiltonian in memory and must only perform a distributed multiplication of the Hamiltonian and wavefunction. Each term in the power sequence is achieved by solving sparse linear equation groups using the Intel Parallel Direct Sparse Solver for Clusters[4] that takes advantage of the matrix sparsity. The Intel Parallel Direct Sparse Solver can solve hundreds of millions of linear equations, which allows our code to handle structures with millions of atoms.

The electronic structure is calculated by the nearest neighbor orthogonal sp$^3$d$^5$s* tight-binding model[5] using literature values for the silicon tight binding parameters.[6] For hydrogen atoms on the surface, we take $ss\sigma = -5.31$ eV and $sp\sigma = 5.50$ eV with $E_H = -3.59$ eV for H-Si bonds. These result from a nonlinear least square fit based on the Levenberg-Marquardt method with finite differentiation on the hybrid functional HSE06 band structures of Si(111) and Si(100) slabs.

The local density of states is extracted from the tight binding wave functions along a high-symmetry sampling line. In the cubic system using fractional coordinates, this line starts from the metabond, passes through the tetrahedral site $(1/4, 1/4, 1/4)$ and octahedral site $(1/2, 1/2, 1/2)$, then stops after another tetrahedral site $(3/4, 3/4, 3/4)$. All the states in a cylindrical disk of 10 Å radius along the above path are counted towards the local density of states at that location. Due to symmetry, this sampling is enough to represent the local density of states over all significant regions of the metalattice.

**Impact of Grain Boundaries on Confinement**

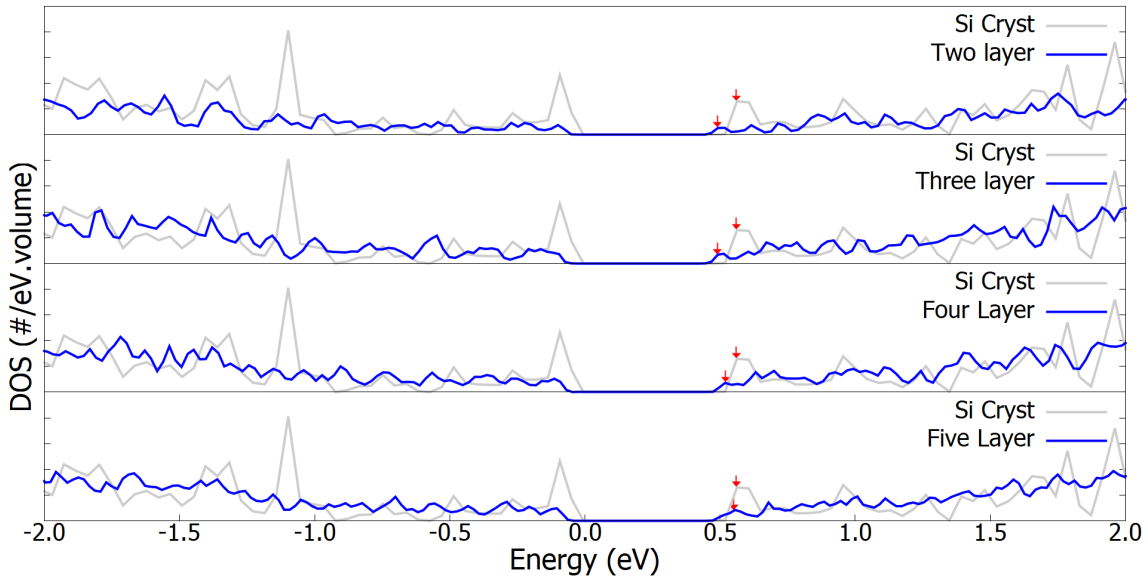

**Figure S3.** Impact of Sigma3 twin grain boundary on conduction band density of states for periodically stacked layers containing such boundaries. Each unit cell of such a slab contains two types of grains rotated by π/6 with respect to each another. The number of layers refers to the thickness of each grain (i.e. the number of Si (111) layers in the normal direction). If the grain size is not too small, the grain boundaries have a minor effect on the location of the conduction band edge.